\documentstyle[12pt]{article}
\begin{document}

\def\simgt{\mathrel{\lower .5ex \rlap{$\sim$}\raise .5ex \hbox{$>$}}}

\begin{titlepage}
\vspace*{1cm}
\begin{center}
{\Large \bf {Universal Scaling of Strong-Field Localization in an Integer
Quantum Hall Liquid}}

\vspace*{0.5cm}
{\bf Dongzi Liu} and {\bf S. Das Sarma\\}

\vspace*{0.5cm}
{\em Center for Superconductivity Research\\
Department of Physics\\
University of Maryland\\
College Park, MD 20742\\}

\vspace*{1cm}

\end{center}

\begin{abstract}


	We study the Landau level localization and scaling properties
of a disordered
two-dimensional electron gas in the presence of a strong external
magnetic field. The impurities are treated as random distributed
scattering centers with parameterized potentials.
Using a transfer matrix for a finite-width strip geometry, we
calculate the localization length as a function of system size and
electron energy.
The finite-size localization length is determined by
calculating the Lyapunov exponents of the transfer matrix.
A detailed finite-size scaling analysis is used
to study the critical behavior near the center of the
Landau bands. The influence of varying the impurity concentration,
the scattering potential range and its nature,
 and the Landau level index on the scaling behavior and on the critical
exponent is systematically investigated.
Particular emphasis is put on studying the effects of finite range of
the disorder potential and Landau level coupling on the quantum localization
behavior.
 Our numerical results, which are carried out on systems much larger
than those studied before, indicate that pure $\delta$-function
disorder in the absence of any Landau level coupling gives rise to
non-universal localization properties with the
critical exponents in the lowest two Landau levels being substantially
different. Inclusion of a finite potential range and/or Landau level
mixing may be essential in producing universality in the localization.

\end{abstract}
\end{titlepage}


\begin{center}
{\Large\bf I. Introduction}
\end{center}

	The quantum Hall effect [1] phenomenon is now well
accepted to be closely related to the strong-field Landau level
localization problem in a disordered  two-dimensional(2D) electron
system. In a disordered 2D system, without any external
magnetic fields, it is well-known that all electronic
states are localized [2] and strictly there is no
metal-insulator transition.
In the presence of a strong external magnetic field
the energy spectrum of a 2D system is a set of impurity-broadened Landau bands.
While the electronic states in the tails of the Landau bands remain
localized, the states at Landau band centers
delocalize and become extended with a mobility edge $E_c$ near the
center of each Landau band. When the Fermi energy is in the
localized regime, the Hall resistance($\rho_{xy}$) is quantized with
vanishing dissipative or longitudinal resistance($\rho_{xx}=0$) at
$T=0$.  When the Fermi energy
moves to the delocalized regime, there is a metal-insulator transition
at $E_c$, and the system behaves as a metal with non-zero $\rho_{xx}$
and non-quantized $\rho_{xy}$ which takes us from one quantized Hall
plateau to the next.
This transition regime, where $\rho_{xx}$ is non-zero and $\rho_{xy}$
is jumping from one plateau to the next, plays an important
fundamental role in localization and quantum Hall studies.
Experimentally [3-9], this transition regime becomes sharper and narrower as
$T$ decreases with the $T=0$ limit thought to be infinitely sharp with
only the state at $E=E_c$ being extended and all other states
localized.
Existence of these extended states(a set of measure zero at $T=0$)
near the Landau level centers is of crucial importance to the quantum
Hall effect phenomenon and, conversely, the quantum Hall effect
demonstrates the existence of extended states (and, of metal-insulator
transitions) in the strong-field 2D system. In this paper, we provide
a detailed numerical study of the localization properties of the two
lowest Landau levels in the strong-field 2D system using a finite-size
scaling analysis. There exist several recent studies of this type in
the literature [10-13]. Utilizing the advantages provided by massively
parallel computing machines, we have been able to go to system sizes
substantially larger (by a factor of four) than those hitherto
existing in the literature. In addition, our emphasis, in contrast to
the earlier numerical studies [10,11] which mostly concentrate on the lowest
Landau level in the presence of zero-range disorder potential, is on
the effects of finite range disorder potentials and Landau level
coupling effects on the localization properties of the two lowest
Landau levels. We are particularly interested in the issue of
universality ({\it i.e.} whether the critical localization  exponents
are always the same independent of Landau level index, type of
disorder potential, {\it etc.}) in strong-field Landau level
localization , a topic which has created some controversy both in the
experimental and in the theoretical literature on the subject [3-14].

	The critical localization behavior of the quantum Hall effect can be
summarized as follows. The quantum Hall resistance $\rho_{xy}$ and
the longitudinal dissipative resistance
$\rho_{xx}$ of a strong-field 2D disordered system are measured at different
temperatures($T$) in the transition regime by
varying the magnetic field($B$).
In the transition regime
it is found [3,4]
that the maxima of $d^{n}\rho_{xy}/dB^{n}$ diverge as
 $T^{-n\kappa}$($n=1,2,3$)  and the width of $\rho_{xx}$
peak, $\Delta B$,  is proportional to $T^{\kappa}$.
Here, $\Delta B$ is the width of the transition regime where
$\rho_{xx}$ is non-zero and $\rho_{xy}$ is unquantized.
  The exponent $\kappa$ of the temperature dependence arises from a
competition between two microscopic length scales at a finite $T$ [14].
One is  the inelastic scattering or the phase breaking
length $L_{in}$ which scales with
temperature with an exponent $p$($L_{in}(T)\propto T^{-p/2}$). The other is the
localization length $\xi$ for the electronic state at the Fermi
energy, which scales with energy with an exponent $\nu$.
The localization length diverges as the Fermi level approaches the
mobility edge critical energy
$E_{c}$ in the middle of each Landau band {\em i.e.}  $\xi (E)\propto \mid
E-E_{c}\mid ^{-\nu}$. When $E_F$ is at the tails of Landau bands,
$\xi\ll L_{in}$, the system is insulating and
localization plays the major role in producing  quantum Hall plateaus.
When $E_F$ is near the
critical energy, $\xi\gg L_{in}$, inelastic scattering length acts as
a cut-off making the
system behave metallic.
Using scaling arguments [14] it is easy to combine these two exponents to
show that the temperature dependence of $\rho_{xx}$ and $\rho_{xy}$ in
the transition regime({\it i.e.} in the metallic phase with $E_F\sim
E_c$) would be controlled by a composite exponent $\kappa=p/2\nu$, which
is directly measured experimentally by studying the $\rho_{xx}$ peak
and the associated $d\rho_{xy}/dB$.
(We have, in fact, recently shown [15] that the scaling analysis can be
extended even to the  $\rho_{xx}$ minima,{\it i.e.} where  $\rho_{xy}$
is quantized, provided the conduction mechanism is activated transport
and not variable range hopping.) A key question in this problem is
whether the exponent $\kappa$ is universal or not. Earlier
experimental investigation [3,4] involving studies of the temperature
dependence of  $\rho_{xx}$ peak and the associated  $d\rho_{xy}/dB$
was controversial with $\kappa$ being independent of Landau level
index in one study and dependent (by a factor of two) on the Landau
level index in the other. There are also some reports [5] of the
dependence of the localization exponent on the impurity scattering
strength. In our opinion, the most thorough and reliable temperature
dependent experiments tend to indicate a universal (Landau level
independent) $\kappa$.
The problem with experimental measurements of $\kappa$ is that it is a
composite scaling exponent, $\kappa=p/2\nu$, being dependent on the
localization exponent $\nu$ and the phase breaking exponent $p$. While
calculations, such as the one carried out in this paper, can provide a
reasonably accurate numerical value of $\nu$(for non-interacting
electrons), the value of the inelastic scattering exponent $p$ in the
strong-field situations is simply not known. Comparison between theory
and temperature-dependent localization experiments has been carried
out mainly under the assumption that $p\approx 2$ which is the
corresponding zero-field clean limit result for electron-electron
scattering in two dimensional Fermi liquids [10]. (We mention that under
the assumption $p=2$, one gets $\kappa\approx 0.4$ using the
numerically calculated $\nu_0\approx 2.3$, obtaining excellent
agreement between theory and experiment.)
There are, however, some
difficulties with this ``standard'' theoretical analysis of the
localization data. First, we do not see any justification for applying
the zero-filed theoretical result for $p$ to the strong-field
situation. Second, there is no justification for using the clean limit
result($p\approx 2$) rather than the disordered 2D result($p\approx
1$) except perhaps the(fortuitous?) ``agreement'' obtained between
theory and experiment using $p=2$. Third, even if $\nu$ is
universal({\it i.e.} Landau level independent, {\it etc.}), it is very
hard to see how $p$ could be universal -- for example, in the
zero-field situation $p=1(2)$ depending  on whether the 2D system is
dirty(clean). Because universality of $\kappa$ requires universality
of both $p$ and $\nu$, we believe that a complete understanding of the
experimental results awaits a theory for inelastic scattering in the
strong-field 2D system.
In this respect, we note that temperature dependent experiments
measuring the width $\Delta B$ of $\rho_{xx}$ peaks and of the
associated $d\rho_{xy}/dB$ show that $\kappa$ in GaAs samples[9]
achieve the ``universal'' value ($\sim 0.4$) only for $T<200mK$ whereas
earlier similar experiments[3] on InP heterostructures produce the
universal $\kappa\sim 0.4$ for $T\sim 4K$. Thus, the size of the
critical scaling regime (which is determined by an interplay between
localization and inelastic scattering) is clearly sample-dependent.
There are also systematic studies of Si-MOSFETs[4,6,7] and GaAs
heterostructures[5] in the literature which find that the measured
value of $\kappa$ is non-universal, with the reported dependence of
$\kappa$ on both the Landau level index and sample quality. A recent
size-dependent experimental localization study[8] of narrow GaAs
heterostructures, which claims separate experimental measurements of
$\nu$ and $p$, concludes that while $\nu(\simeq 2.3)$ is universal
(independent of both Landau level index and the sample quality), $p$
varies from 2.7 to 3.4, depending on the Landau level index and the
sample quality. Thus, $\kappa$ itself($\kappa=p/2\nu$) may not be
universal. We emphasize that our work is solely on an accurate
numerical calculation of the localization exponent $\nu$ and the
universality being discussed in this paper refers entirely to the
localization aspect of the problem({\it i.e.} we are concerned with
whether or not $\nu$ is independent of Landau level index, the nature
of the impurity scattering potential, {\it etc.}). We have nothing to
say about the exponent
$p$ which is often needed in understanding the experimental
quantum Hall localization data.

The interesting recent experimental attempt [8], alluded to above,
circumvent the problem of having a composite exponent $\kappa$ by
direct and independent measurement of $\nu$(and $p$) using
experimental 2D strip systems of various widths warrants a discussion.
 The competing
length scales in this type of experimental ``finite-size scaling''
studies are the localization length and the system size (with the
inelastic scattering length, which is not playing an active role,
assumed to be a constant, larger than the system size,
 because the temperature is held fixed in
these experiments). Such direct measurements of $\nu$ show that it is
indeed universal with a value $\nu\approx 2.3$ in excellent agreement
with the existing (and our own) theoretical calculations. While the
basic premise of these experiments is novel and this procedure
certainly eliminates the difficulty associated with $\kappa(=p/2\nu)$
being a composite exponent, we feel that there are some inherent
problems in the interpretation of these experimental results as well.
Suppose the temperature is such that $L_s>L_{in}$, where $L_s$ is the
finite size of the experimental system. In that situation, obviously
$L_s$ is irrelevant for the localization behavior of the system and
the competition between the localization length $\xi$ and the
inelastic scattering length $L_{in}$ controls the scaling behavior.
Thus, for the finite-size scaling experiments to be meaningful, one
must have $L_{in}>>L_s$, which is precisely the mesoscopic conductance
fluctuation regime [16] where the system is manifestly non-self-averaging
and one should see quantum interference induced conductance
fluctuations because the system size is smaller than the phase
coherence length. Such strong mesoscopic fluctuations are not present in the
experimental results and, therefore, it is unclear to what extent the
necessary condition $L_{in}>>L_s$ is satisfied. We emphasize that,
while the weak-field conductance fluctuation phenomenon is currently
well understood, there has been very little work on strong-field
conductance fluctuation phenomenon, and, in principle, it is
possible that there is a very good (but presently unknown) reason why
conductance fluctuations are absent in the strong-field 2D systems
even in the mesoscopic $L_{in}>>L_s$ regime. Without such a theory
explaining the absence of strong-field mesoscopic conductance
fluctuations, we feel that a complete understanding of the novel
finite-size scaling experiments remains elusive.

	In this paper,
	we study  localization and scaling properties of a disordered
two-dimensional system of finite width in the presence of a strong
magnetic field. The impurities are treated as random distributed
scattering potential centers. By choosing an appropriate cut-off in
the scattering
range, we can divide our system into cells with only the nearest-neighbor
intercell scattering being appreciable. This allows us to set up an intercell
transfer matrix for evaluating the exact single particle
non-interacting electron wavefunction in the presence of disorder.
The finite-size localization length of the electron is determined by
calculating the Lyapunov exponents of the transfer matrix on a
massive parallel processing machine.
Finite-size scaling method
is applied to study the critical behavior near the center of the
Landau bands. The influence of varying the impurity concentration,
the scattering potential and the Landau level index on the critical
exponent is systematically investigated. A short report of our work
has earlier appeared in the literature[17].

     The rest of the paper is organized in the following way: in Sec.
II we give the details of our theoretical method with the relevant
formulas and equations; in Sec. III we present our numerical results;
we provide a discussion with the conclusion in Sec. IV.


\vspace{0.5cm}
\begin{center}
{\Large\bf II. Theoretical Method}
\end{center}

	For a non-interacting 2D electron gas in the presence of a
strong external magnetic field ${\bf B}\equiv(0,0,B)$
perpendicular to the 2D $x-y$ plane,
the single electron energy spectrum is  given by a set of discrete
Landau levels (denoted by the index $N=0,1,2,...$):
\begin{equation}
 E_{N}=(N+\frac{1}{2})\hbar\omega_{c},
\end{equation}
where $\omega_{c}=eB/mc$ is the cyclotron frequency, and $m$ the electron
effective mass. Each Landau level is highly degenerate with a
macroscopic degeneracy given by $(2\pi l_{c}^{2})^{-1}$ per unit area
where $l_{c}=(\hbar c/eB)^{1/2}$ is the magnetic length. By choosing
the Landau gauge for the vector potential ${\bf A}=(0,Bx,0)$, and periodic
boundary conditions in the $y$ direction, the electron wavefunction
becomes a simple harmonic oscillator eigenfunction  in the $x$
direction, and a free-electron plane wave in the
$y$ direction. The one electron Landau wavefunction of the pure system
can then be written as
\begin{equation}
 \phi_{Nk}=\frac{1}{\sqrt{M}}e^{-iky}H_{n}(\frac{x-kl_{c}^{2}}{l_{c}})
    e^{-\frac{1}{2}(\frac{x-kl_{c}^{2}}{l_{c}})^{2}}.
\end{equation}
Here $M$ is the width of the 2D system in $y$ direction,
$H_{n}$ is the Hermite polynomial,  and $k=\frac{2\pi}{M}n$ where
$n$ is an integer. Note that the simple harmonic oscillator has a
displaced center with the oscillation center at $x\equiv
X=kl_{c}^{2}$.
We choose these one electron Landau
eigenstates as a complete basis in the Hilbert space for our
localization study of the
disordered 2D system.

     We consider a two-dimensional disordered system of length $L$
(along the $x$ direction) and width
$M$ in the strip geometry with periodic boundary conditions applied
in the $y$ direction.
We use
$l_{o}=\sqrt{2\pi}l_{c}$ as the length unit in
this problem. The single electron
Hamiltonian in the presence of disorder can be written as
\begin{equation}
 H=\sum_{NX}\mid NX>(N+\frac{1}{2})\hbar\omega_{c}<NX\mid \nonumber\\
    +\sum_{NX}\sum_{N'X'}\mid NX><NX\mid V\mid N'X'><N'X'\mid
\end{equation}
where $\mid NX>$ is the Landau state for the $N$th Landau level with
the center of
oscillation $X$ taking discrete values with spacing $1/M$ due to the
periodic boundary conditions.
The disorder potential $V$, which is treated as the one electron
potential arising from a  random
distribution of impurity  scattering centers, is given by
\begin{equation}
  V({\bf r})\equiv\sum_{i}V_{i}({\bf r}-{\bf r}_{i})
\end{equation}
where ${\bf r}_{i}$ is the location of the $i$th impurity scattering
center and $V_{i}$ depends on
the type of scattering potential.
 We consider two models for the random disorder potential, namely, the
short-range potential , $V_{i}=V_{o}\delta({\bf r}-{\bf r}_i)$, and,
the finite range
Gaussian potential, $V_{i}=\frac{V_{o}}{\pi d^{2}}e^{-|{\bf
r}-{\bf r}_i|^{2}/d^{2}}$. The potentials are chosen to be all attractive,
or all repulsive,  or an equal mixture  of both.
 The concentration of impurities, which along with $V_o$ determines
the scattering strength, is
denoted by $c_{i}$.
An important numerical quantity in our calculation is $l_{cutoff}$
which determines the length scale above which the impurity scattering
potential does not connect the different Landau states, {\it i.e.} we
only need to consider scattering between states $X$,$X'$ satisfying
$\mid X-X'\mid\leq
l_{cutoff}$. Obviously $l_{cutoff}$ is determined by the
characteristics of the Landau wavefunctions and the impurity potential
and concentration.
Our criterion to determine $l_{cutoff}$ is that the
matrix element $<NX\mid V\mid N'X'>$ for $\mid X-X'\mid >l_{cutoff}$
should be much smaller (less than $1\%$)
 than that for $\mid X-X'\mid\leq l_{cutoff}$.
  In our calculations,
we choose $l_{cutoff}=1$ for the $N=0$ Landau level, and
$l_{cutoff}=2$ for higher Landau levels($N=1$). Our choice of
$l_{cutoff}$ is consistent with the others existing in the literature [10,11].
In addition to the inter-Landau-level energy separation
$\hbar\omega_{c}$, there is
another energy scale in this problem which is the Landau level
broadening $\Gamma_{N}$ determined by the strength of the
impurity-scattering potential. Within the
self-consistent-Born-approximation(SCBA), and in the strong field limit
({\it i.e.} no Landau level coupling),
the density of states(DOS) for the short-range impurity potential
is given by [18]:
\begin{equation}
 D(E)=2\sum_{N=0}(\pi\Gamma)^{-1}\left[ 1-\left[\frac{E-E_{N}}
 {\Gamma}\right]^{2}\right]^{1/2},
\end{equation}
where $\Gamma=2V_{o}\sqrt{c_{i}}\ll\hbar\omega_{c}$.
Defining $\gamma=\Gamma/\hbar\omega_{c}$, we note that $\gamma<<1$
defines  the
weak disorder strong-field limit when scattering to other Landau
levels may be ignored.
In some situations we would consider
the strong disorder case, $\gamma\sim 1$, where we have to take into
account Landau level
coupling in our calculations. We choose $\hbar\omega_{c}$ as the unit
of energy in our calculations (which is the most obvious natural choice).

Our localization calculation proceeds in two steps: First, we obtain
the localization length for a 2D strip of a finite width $M$ and an
essentially infinite length of $L=10^5(>>M=4-256)$ by calculating
the system Lyapunov exponent on a massively parallel processing
machine; then, we carry out a finite size scaling analysis to obtain
the localization length in the $M\rightarrow\infty$ limit.
In order to calculate the localization length for a 2D strip of length
$L$ and width $M$,  we start by
dividing the system into cells with length $l_{cutoff}$ and width $M$
such that the next-nearest-neighbor intercell interaction can be
neglected. The number of states in each cell for each Landau level is
$n=l_{cutoff}M$. Expanding the
wavefunction $|\Psi>$ of the disordered system for energy $E$ in the
complete basis of Landau wavefunctions. we get
\begin{equation}
 \mid\Psi>=\sum_{Ni}a_{Ni}\mid NX_{i}>
\end{equation}
with
\begin{equation}
 H\mid\Psi>=E\mid\Psi>.   \label{eq:shro}
\end{equation}
Combining Eqs.(1),(3),(6) and (7) we get
\begin{equation}
 (N+\frac{1}{2})\hbar\omega_{c}a_{N,i+n} + \sum_{N'}
\sum_{k=0}^{2n}<NX_{i+n}\mid
V\mid N'X_{i+k}>a_{N',i+k} = Ea_{N,i+n}.   \label{eq:coe}
\end{equation}
Defining column vectors
\begin{equation}
 A_{i}=\left( \begin{array}{l}
         a_{0i}\\a_{1i}\\ \vdots \\a_{N_{L}-1,i} \end{array} \right),
\end{equation}
and
\begin{equation}
 \psi^{(i)}=\left( \begin{array}{l}
         A_{i+2n-1}\\ \vdots \\A_{i} \end{array} \right),
\end{equation}
where $N_{L}$ is the number of Landau levels being included
in our numerical calculation,
Eq.(\ref{eq:coe}) can be transformed to the following set of linear equations:
\begin{equation}
 \sum_{k=0}^{2n}H_{i+k}A_{i+k}=EA_{i+n},    \label{eq:come}
\end{equation}
where
\begin{equation}
\begin{array}{l}
   H_{i+k}= \left( \begin{array}{cccc}
         \frac{1}{2}\hbar\omega_{c} & 0 & ... & 0  \\
         0& \frac{3}{2}\hbar\omega_{c} & ... & 0  \\
  \vdots &\vdots &\ddots &\vdots  \\
    0 &0 & ... &
  (N_{L}-\frac{1}{2})\hbar\omega_{c} \end{array} \right)
 \delta_{kn} \nonumber \\
  +  \left( \begin{array}{ccc}
        <0X_{i+n}\mid V\mid 0X_{i+k}> &
  ... & <0X_{i+n}\mid V\mid N_{L}-1,X_{i+k}>  \\
         \vdots &\ddots &\vdots  \\
    <N_{L}-1,X_{i+n}\mid V\mid 0X_{i+k}> & ... &
  <N_{L}-1,X_{i+n}\mid V\mid N_{L}-1,X_{i+k}> \end{array} \right).
\end{array}
\end{equation}
Equation (\ref{eq:come}) shows that $A_{i+2n}$
is determined by $A_{i},A_{i+1}...A_{i+2n-1}$
{\em i.e.} we can set up the transfer matrix $T^{(i)}$ such that
\begin{equation}
   \psi^{(i+1)}=T^{(i)}\psi^{(i)}.
\end{equation}
Here the transfer matrix $T^{(i)}$ is a $(2n\times N_{L})\times
(2n\times N_{L})$ dimensional matrix:
\begin{equation}
  T^{(i)}=\left( \begin{array}{cccccc}
    -H_{i+2n}^{-1}H_{i+2n-1} & ... &  -H_{i+2n}^{-1}(EI-H_{i+2n})&
   ... & -H_{i+2n}^{-1}H_{i+1} & -H_{i+2n}^{-1}H_{i}\\
    I &... & 0 &... &0 & 0 \\
   \vdots & \ddots &\vdots &\ddots &\vdots &\vdots \\
   0 &... &I &... &0 &0 \\
   \vdots & \ddots &\vdots &\ddots &\vdots &\vdots \\
   0 &... & 0 &...& I & 0 \end{array} \right)
\end{equation}
where $I$ is $N_{L}\times N_{L}$ unit matrix.
Note that if Landau level coupling is neglected in the calculations
(as has been done in all the existing calculations), $N_L=1$.
We have done calculations for single Landau levels ($N=0,1$) with
$N_L=1$ and for coupling
between the lowest two Landau levels with $N_L=2$. $T^{(i)}$ is
calculated by summing up the
contribution from each impurity scatterer, and, therefore, depends on the
details
of the local distribution of the random impurities. For example,
for the short-range impurity potential $V_{i}=V_{o}\delta ({\bf r}-{\bf
r}_{i})$, we obtain the following non-zero matrix elements:
\begin{eqnarray}
  <0X_{n_{1}}\mid V_{i}\mid 0X_{n_{2}}>&=&\frac{\sqrt{2}V_{o}}{M}e^{i2\pi
 (n_{1}-n_{2})y_{i}/M}e^{-\pi [(x_{i}-n_{1}/M)^{2}+(x_{i}-n_{2}/M)^{2}]} ,\\
 <0X_{n_{1}}\mid V_{i}\mid 1X_{n_{2}}>&=&
\frac{\sqrt{2}V_{o}}{M}2\sqrt{\pi}
(x_{i}-n_{2}/M)e^{i2\pi (n_{1}-n_{2})y_{i}/M}
\nonumber \\  & &\times
e^{-\pi\ [(x_{i}-n_{1}/M)^{2}+(x_{i}-n_{2}/M)^{2}]}, \\
 <1X_{n_{1}}\mid V_{i}\mid 0X_{n_{2}}>&=&\frac{\sqrt{2}V_{o}}{M}
2\sqrt{\pi}(x_{i}-n_{1}/M)
e^{i2\pi
 (n_{1}-n_{2})y_{i}/M}
\nonumber \\
 & &\times e^{-\pi [(x_{i}-n_{1}/M)^{2}+(x_{i}-n_{2}/M)^{2}]}, \\
  <0X_{n_{1}}\mid V_{i}\mid 0X_{n_{2}}>&=&\frac{\sqrt{2}V_{o}}{M}
4\pi (x_{i}-n_{1}/M)(x_{i}-n_{2}/M)e^{i2\pi (n_{1}-n_{2})y_{i}/M}
\nonumber \\  & &\times
e^{-\pi [(x_{i}-n_{1}/M)^{2}+(x_{i}-n_{2}/M)^{2}]}.
\end{eqnarray}
We do not show here the explicit matrix elements for the Gaussian
impurity potential which can also be calculated analytically.

As mentioned before, once the transfer matrix $T^{(i)}$ is formed,
we calculate the localization length by
evaluating the Lyapunov exponent for the
transfer matrix.
Calculation of the Lyapunov exponent for maps such as the one defined
in Eq.(13) is standard in non-linear dynamics, and has been
successfully used in localization calculations in other contexts [19].
 For the transformation $T$ which transforms a vector $\psi$
according to Eq.(13), the
Lyapunov exponents are given by
\begin{equation}
  L_{y}=\lim_{k\rightarrow\infty}\frac{1}{k}\ln
     \frac{\mid\psi^{(k+1)}\mid}{\mid\psi^{(1)}\mid}
=\lim_{k\rightarrow\infty}\frac{1}{k}\ln
     \frac{\mid\prod_{i=1}^{k} T^{(i)}\psi^{(1)}\mid}{\mid\psi^{(1)}\mid}.
\end{equation}
For our ($L,M$) strip geometry, with $L$ very large and $M$ small,
 $\mid\psi^{(1)}\mid\sim\mid\psi(0)
\mid$, $\mid\psi^{(k+1)}\mid\sim\mid\psi(L)\mid$,
$k=\frac{L}{\Delta X}=LM$. For localized states, the localization length
$\lambda_{M}$ is defined to be the characteristic length controlling
exponential decay of the wavefunction,
{\em i.e.} $\mid\psi(L)\mid
/\mid\psi(0)\mid\sim e^{-L/\lambda_{M}}$, so that
\begin{equation}
     \lambda_{M}=-\frac{1}{ML_{y}}.
\end{equation}
Mathematically there exist $2n\times N_{L}$ Lyapunov exponents for
the transfer matrix $T$, but the relevant one for the localization
length is the smallest negative Lyapunov exponent.

We employ the following technique to calculate the smallest negative
Lyapunov exponent.
Denote $S$ as an upper triangular matrix, {\em i.e.}
$S_{i>j}=0$, $Q$ as a lower triangular matrix, {\em i.e.} $Q_{i<j}=0$
with $\det Q=1$. $T^{(1)}$ can be expressed as
$T^{(1)}=Q^{(1)}S^{(1)}$, then $T^{(2)}Q^{(1)}=Q^{(2)}S^{(2)}$ and so
on. Finally,
\begin{equation}
 \prod_{i-1}^{k} T^{(i)}=Q^{(k)}\prod_{i=1}^{k} S^{(i)}
\end{equation}
The relevant Lyapunov exponent we need is
\begin{equation}
    L_{y}=\lim_{k\rightarrow\infty}\frac{1}{k}\ln\mid\prod_{i=1}^{k}
S^{(i)}\mid_{n\times N_{L}+1, n\times N_{L}+1}
\end{equation}
The problem of computing the Lyapunov exponent now reduces to a simple
matrix multiplication problem. The problem is, therefore, ideally
suited for the parallel processing capability of massive parallel
processing computers. We have used both CM2 and CM5 parallel
processing computers of the Thinking Machine Co. to carry out our
calculations. Note that, since scattering from various impurities is
independent of each other, we can use a parallel processing algorithm to
calculate their contribution to $T$ as well.

Once the localization length, $\lambda_M(E)$, at a particular energy
$E$ and for a specific finite value of $M$ is obtained from the
Lyapunov exponent calculation, we carry out a finite-size scaling
analysis to go to the 2D limit({\it i.e.} the $M\rightarrow\infty$
limit). Finite-size scaling analysis has been extensively used in
localization calculations and for details we refer to the literature
[10,11].
The idea is that one should see
one-parameter scaling for large enough sample width $M\geq
M_{sc}$ and for energies in the  critical regime $\mid E-E_{c}\mid\leq
E_{sc}$(note that the system length $L$ is very large).
\begin{equation}
    \frac{\lambda_{M}(E)}{M} = f\left(\frac{M}{\xi (E)}\right).
\end{equation}
Here $f(x)$ is the  universal scaling function for the localization
problem, with $\xi(E)$ the localization length for the infinite
system. This means that in the critical
regime a  change in energy is equivalent to a  change in the system size.
As we go to large system sizes, the localized states far away from
the critical region have $\xi (E)\ll M$, and
$\lambda_{M}\sim \xi$, so the scaling function
has the asymptotic behavior of $f(x)\sim 1/x (x \gg 1)$. On the other hand, for
the states very close to the critical energy $E\sim E_c$, $\lambda_{M}
\sim M$, and
$\xi\gg M$, so that  $f(x)\sim$constant for $x\ll 1$.
By plotting the numerically calculated $\lambda_M(E)/M$ against $M$ on
a log-log plot, we can calculate $\xi(E)$ by using a least-square
optimization fit to make all the data points collapse on a single
``smooth'' scaling curve for $\lambda_M(E)/M$ against $M/\xi(E)$
obeying Eq.(23) and the correct asymptotic properties. This
finite-size scaling analysis immediately gives us $\xi(E)$, the
localization length for the infinite system.
Using log-log plot of the calculated $\xi (E)$ against $E$, we can
obtain  the critical
exponent $\nu$, {\em i.e.}
\begin{equation}
 \xi (E)\propto\mid E-E_{c}\mid ^{-\nu}.
\end{equation}
Obeying of both Eq.(23) and (24) together puts rather stringent
requirements on the finite-size scaling analysis ensuring the correct
evaluation of the critical exponent $\nu$ from the optimization
procedure. Note that the long strip-geometry topology of our model
ensures self-averaging for the Lyapunov exponent calculation of the
transfer matrix and no additional averaging is required.


\vspace{0.5cm}
\begin{center}
{\Large\bf III. Numerical Results}
\end{center}

 In this section we present detailed numerical results obtained by
applying the theoretical method described in II. We have
carried out calculations on system sizes up to $M=256$, which is larger than
 those studied before [10,11]. The sample length $L(\sim 10^5)$ is chosen such
that
the numerical error in the calculation of
Lyapunov exponent is less than $1\% $.
We consider a wide range of impurity
concentrations from $c_{i}=2$ to $c_{i}=32$. We  consider
 the short-range $\delta$-function scattering  potential and
Gaussian long-range potentials with the potential range
$d/l_{c}=0.5$ and $1$.
For the weak disorder case, we carry out calculations
for single({\it i.e.} $N_L=1$) Landau levels $N=0,1$, and  for the
strong disorder case, we take into account  Landau level
coupling($N_L=2$) between $N=0$ and $1$ Landau levels.

 In Figs.1 and 2 we present our numerical results for Landau levels
$N=0$ and 1 respectively in the weak disorder({\it i.e.} neglecting
Landau level coupling), symmetric({\it i.e.}
with equal number of
repulsive and attractive $\delta$-function scatterers) case.
 The weak disorder results for different
impurity concentrations with symmetric $\delta$-function scatterers
are summarized in Table 1.
The calculated localization critical exponent $\nu$
depends strongly on the Landau level index,  but varies  little with the
impurity concentration.
This finding of a strong variation of $\nu$ with the Landau level
index $N$ is consistent with earlier findings in the literature [10,11](using
substantially smaller system sizes).
While our numerical results show reasonably
good one-parameter scaling (Fig.1(b) and Fig.2(b))
properties for both the Landau levels, the asymptotic value of the
scaling function seems to  depend
on the  Landau level index. In our scaling plots, the saturation
values for $M/\xi\ll 1$ are quite
different for the different Landau levels (for $N=0$,
$\lambda_{M}/M\rightarrow 1$; for $N=1$, $\lambda_{M}/M\sim 2$).
The insensitivity of the scaling function to a variation in the
impurity concentration $c_i$ is, however, quite robust.

For random distributed weak-disorder repulsive $\delta $ impurities,
the calculated critical
exponents are presented in Table 2 and a typical set of scaling data is shown
in
Fig.3. In this asymmetric situation with all repulsive scatterers,
 the critical energy $E_c$ is shifted away from the exact
center of the Landau band, but as the impurities get denser,
the critical energy approaches the Landau band center. In the
dilute impurity limit, the critical exponents are asymmetric, being
somewhat different  for the lower
and the higher energy branches of the Landau band. It is known
that in the low impurity concentration case, the density of states
(DOS) is asymmetric with the  maxima of DOS shifting away from the center of
the Landau band (for repulsive scatterers, the maxima shift to lower
energies) [20].
We believe the asymmetry in our calculated critical exponents arises from
this finite impurity concentration effect, and should disappear if
Landau level coupling is included in the calculation.

 Our results for the values of $\nu$ in the presence of Gaussian
impurity scattering potential (without any Landau level coupling) is
summarized  in Table 3, and some typical scaling plots
are shown in Fig.4 (for $N=0$) and Fig.5 (for
$N=1$).
Basic trends are similar to those found in the $\delta$-function case
with one important difference.
While for the $N=0$ Landau level, the
critical exponent remains essentially a constant
($\nu\simeq 2.2$) as the potential range is varied, the exponent
for the $N=1$ Landau level clearly decreases with
increasing the range of the
impurity potential.
In particular, the value of $\nu$ decreases from around 5.5 for
zero-range potential to around 2.5 for finite range potentials (the
value of $\nu_0$, however, remains constant around 2.2). Based on this
rather limited numerical data, one may speculate that universality is
restored by finite-range scatterers and that $\delta$-function
scattering is quite a special case.

 In the presence of strong disorder, the Landau level broadening
$\Gamma $ is comparable to inter-Landau-level energy difference
$\hbar\omega_{c}$, and, therefore, we must include Landau level
coupling in the calculation.
We consider the simplest situation taking into account the
coupling between $N=0$ and $N=1$ Landau
levels. The strength of the Landau level coupling is parameterized
 by $\gamma =\Gamma/\hbar\omega_{c}$.
Our calculated
numerical results for the critical exponents(using the
$\delta$-function potential model) are summarized in Table 4.
In Fig.6 we show one set of representative scaling data for the finite
$\gamma$ Landau level-coupled case. We observe that the calculated critical
exponents for the lower and the higher energy branches of the same Landau band
are
different due to the strong asymmetry introduced by Landau level
coupling effect($E_c\neq 0$).
This asymmetry is introduced by our model which considers
coupling only between the lowest two Landau levels. The
coupling obviously plays a much stronger role in the higher(lower)
energy branch of the $N=0(1)$ Landau
band because the coupling is the strongest among those states. The
most important result of our Landau level coupled calculations is that
$\nu_0$ and $\nu_1$ come closer together in the presence of Landau
level coupling, allowing us to speculate that universality may be
restored {\bf even for the $\delta$-function scattering potential}
once Landau level coupling is included in the theory.

We have also carried out calculations for two coupled Landau levels in
the presence of finite range impurity potentials. These calculations,
which are necessarily on somewhat smaller system sizes, bear out our
basic conclusions that finite potential range and Landau level
coupling bring $\nu_0$ and $\nu_1$ close together numerically.
We emphasize that by necessity our numerical results for the finite
range disorder potential and for the Landau level coupling case are
much less quantitatively accurate than the simple zero-range uncoupled
calculations. The qualitative trends, however, are reliable and can be
trusted. For quantitatively accurate results in the presence of finite
range disorder potential and Landau level coupling effects, one needs
to consider substantially larger values of system width which are not
currently computationally accessible. We refrain from showing error
bars for our finite range disorder and Landau level coupling
calculations except to mention here that the error bars in Tables 2-4
are at least an order of magnitude larger than those in Table 1.


\vspace{0.5cm}
\begin{center}
{\Large\bf IV. Discussions and Conclusion}
\end{center}

In this paper, we study localization and scaling properties of a
disordered
two-dimensional electron system in the presence of a strong external
magnetic field, with emphasis on the question of universality of the
localization transition in various Landau levels.
 The impurities are treated as random distributed scattering centers with
short range $\delta$-function  or finite  range Gaussian
potentials.
Our localization calculation employs a strip geometry for which we
obtain the Lyapunov exponent by setting up a suitable transfer matrix.
This is done rather efficiently in a parallel algorithm.
The Lyapunov exponents of the transfer matrix are calculated to determine
the finite-size localization length.
The critical behavior of the disordered two-dimensional system
near the center of the Landau bands is then studied
using the standard one-parameter finite-size scaling analysis.
 We systematically investigate
the influence of varying the impurity concentration,
the scattering potential, and the Landau level index on the critical
exponent. We also study the effect of Landau level coupling on the
localization exponent.

  In the weak disorder limit, we consider scattering only within a
single Landau level, neglecting any Landau level coupling coupling effect.
 We find that the critical exponent and the
scaling property both depend on the Landau level index, particularly
for the $\delta$-function disorder. Within a
single Landau level,  especially the lowest Landau level, we observe
universal scaling behavior independent of impurity concentration,
scattering potential type (repulsive or attractive or both,
 short range or long range). The calculated
critical exponent varies a little for different impurity
concentrations which we attribute to finite-size or
finite-concentration effect. We believe that the dependence of the
critical exponent on the Landau level index comes from the different
symmetry properties of the Landau wavefunction in different Landau
levels.  For short-range scattering potentials with the range
smaller than or comparable to the size  of the
Landau wavefunction, the detailed nature of the Landau wavefunction
may  play
a role in the critical behavior, leading to different values of
$\nu_N$ for $N=0$ and 1. In the presence of only repulsive
scattering potential, we observe an expected asymmetry in the critical exponent
arising from the asymmetry of the density of states
which is due to the finite impurity concentration effect. For finite
 range scattering
potentials, we see the clear trend of decreasing  critical exponent
with increasing potential range
for the higher
Landau level $N=1$(without much effect on $\nu_0$). We speculate that
for real long range
scattering potentials with range larger than the size of Landau
wavefunctions, the dependence of the critical behavior on the details
of the symmetry of the Landau wavefunction will disappear,
producing true universality in the critical localization behavior.
This percolation limit, beyond the scope of our work, has actually
been studied both analytically and numerically [21-23].

  In the strong disorder limit, the Landau level broadening due to the
presence of disorder is comparable to the inter-Landau-level energy
difference, so the coupling among Landau levels should be taken into
account. It is, in fact, not justified to ignore lower Landau
levels even in the single-Landau-level approximation when one is
considering the localization properties of the excited levels({\it
e.g.} $N=1$).
In our numerical calculation, we just consider coupling between the
lowest two Landau levels. This is, of course, a very crude
approximation for Landau level coupling which we are forced to make
  because of our limited numerical
capability. We find that Landau level coupling, even in this
perturbative sense, has some qualitative effect.
For the $\delta$-function potential
we find that in the strong disorder Landau level coupling regime,
the numerical values of the critical exponents for different Landau
levels get closer as we
increase the impurity concentration while the scaling properties
remain the same as in the non-coupling case. The critical energies shift from
the center of each Landau band because of the asymmetry introduced by
Landau level coupling.
This new critical behavior arises
from the mixing of symmetry of different Landau wavefunctions.
A similar trend of decreasing critical exponent in the
higher($N=1$) Landau level is seen for finite
range Gaussian impurities as well.
We believe that both finite range disorder and Landau level coupling
reduce the role of the detailed symmetry of the Landau wavefunction in
the critical localization problem and restore universality in the
Landau level localization properties. Our limited numerical results
only indicate a trend and much larger simulations are needed to
definitely establish this claim.

 There have been several earlier numerical investigations [10-13] of
strong-field localization in disordered two-dimensional electron
systems. Our parameterization of the random disorder potential (as
arising from randomly distributed point impurity scattering centers
characterized by short-range $\delta$-function or finite-range
Gaussian potentials) is similar to that used by Ando and Aoki [11] who
concentrated mainly on the large impurity concentration
limit($c_i=40$), neglecting Landau level coupling, and used rather
small system sizes (up to $M=16$ only). Ando and Aoki [11] concluded that
$\nu_0\simeq 2$ and $\nu_1\simeq 4$ for $\delta$-function scatterers
compared with our results $\nu_0=2.2\pm 0.1$ and $\nu_1\simeq 5.5\pm
0.5$. For finite-range scatterers, they found $\nu_0\simeq 2$ for
$d=2$, but did not calculate $\nu_1$. We also find that $\nu_0(\simeq
2.2)$ remains unaffected by the finite range of the scattering
potential whereas $\nu_1$ is reduced considerably($\nu_1\leq 3$),
bringing it much closer to $\nu_0$ and thus restoring universality.
The work of Ando and Aoki was criticized (for having used system sizes
far too small to see conclusive evidence of scaling) by Huckestein and
Kramer [10] who went to larger system sizes($M=64$) and concluded that
$\nu_0=2.34\pm 0.04$ for short-range white-noise random potential.
These latter authors [10] did not systematically investigate the exponents
in the higher Landau levels or the effects of having a finite disorder
potential range. They also neglected Landau level coupling effects.
Our $\delta$-function potential numerical results for the lowest
Landau level are quantitatively and qualitatively consistent with the
findings of Huckestein and Kramer. It should be mentioned that
Huckestein and Kramer [10] model the disorder potential somewhat
differently from us -- instead of working with random distributed
impurity centers, they use a random white-noise distribution of
disorder potential matrix elements with a vanishing correlation
length. We point out that the largest system sizes($M=256$) used in
our work are much larger than those used by Huckestein and Kramer. (It
is, in some sense, essential that we use larger systems because we are
interested in investigating higher Landau levels, finite range random
potentials, and Landau level coupling effects.) For slowly varying
random disorder potentials, the strong-field quantum localization
problem is equivalent to a classical percolation problem, which has
been analytically solved [21] to produce $\nu=4/3$. If quantum tunneling is
included in the percolation calculation[23], one gets $\nu=7/3\simeq 2.33$
which is very close to our numerically calculated $\nu_0$. It is
worthwhile to mention that in the percolation limit, the Landau levels
are necessarily coupled and the disorder is necessarily long-range --
it is, therefore, reassuring to know that our finite-range disorder
and Landau level coupled calculations produce $\nu_0\approx\nu_1$
which are consistent with the percolation results. There has also been
a numerical calculation [22] near the percolation threshold, including
effects of quantum interference and tunneling, leading to $\nu=2.5\pm 0.5$.

Existing numerical and analytical calculations [10-13,21-23] clearly establish
that
the localization critical exponent is around 2.3 independent of the
Landau level index for finite-range potentials and/or in the presence
of Landau level coupling, and for the $\delta$-function potential at
least in the lowest Landau level. We emphasize that the percolation
limit necessarily includes Landau level coupling and long-range
disorder, and is thus complementary to the $\delta$-function
short-range disorder in the absence of Landau level coupling. The fact
that both the percolation calculations  and the lowest Landau level
finite size scaling analysis  for $\delta$-potantials gives the same
critical exponent $\nu\approx 2.3$ provides evidence in favor of
universality in the Landau level localization. As mentioned before,
our finite range and/or Landau level coupled numerical calculations
are semi-quantitatively consistent with the percolation results. On
the other hand, there does seem to be a problem with the
$\delta$-function random potentials in the higher($N=1$) Landau
levels(in the absence of any Landau level coupling) where at least
four independent studies [10,11,12,17](including our own, which employs the
largest
system sizes) find the critical exponent $\nu_1$ to be substantially
different from that in the lowest Landau level($\nu_1\approx 5,
\nu_0\approx 2.3$). It is, of course, possible (but highly unlikely in
our opinion) that this lack of universality for the short-range
$\delta$-function disorder is purely a finite-size phenomenon and
future calculations involving larger system sizes will restore
universality ({\it i.e.} $\nu_1=\nu_0$) even for $\delta$-function
disorder. This scenario is unlikely in our view for two reasons: (1)
Early calculations involving small system sizes($M=16$) all the way to
our large system($M=256$, we have some limited results for $M=512$)
calculations consistently show $\nu_1\simgt 2\nu_0$ for
$\delta$-function disorder (in the absence of Landau level coupling)
-- the quality of scaling for $N=1$ Landau level in our calculations is
quite comparable to that for $N=0$ level, and, therefore, it is
unclear how increasing system sizes further would reduce $\nu_1$ by
more that a factor of two to bring it in agreement with $\nu_0$; (2)
for $\delta$-function disorder, we see no obvious intrinsic length
scale in the problem which demands that we go to system sizes
substantially larger than this characteristic length scale to obtain
the {\bf true} (rather than effective) exponents -- thus it is very
hard to understand the nature of any crossover behavior which may be
dominating the finite-size scaling analysis for $N=1$ producing
$\nu_1\approx 5$(we do expect that one has to use somewhat larger sizes
for analyzing the critical behavior of the $N=1$ level compare to the
$N=0$ behavior because the free electron Landau wavefunction for $N=1$
is more spread out than that for the $N=0$ level).
Based on these considerations we feel that ({\bf in the absence of
Landau level coupling}) the range of the disorder potential may
actually be a relevant perturbation, and the localization transition
in higher Landau levels may be non-universal for short-range disorder
as our numerical calculations indicate. We believe that this lack of
universality arises from the fact(Fig.7) that the matrix element of the
disorder potential projected onto a particular Landau level shows
qualitatively different spatial behavior for the higher Landau levels
compared with the lowest Landau level as a function of the range of
disorder. This is explicitly shown in Fig.7.
In particular, the spatial behavior of the lowest Landau
level matrix element remains qualitatively the same (namely, it is a
monotonically decreasing function of distance) independent of the
range of the disorder potential (because the wavefunction for the
lowest Landau level is nodeless). On the other hand, the higher Landau
levels being excited states have nodes and, therefore, the spatial
behavior of disorder matrix elements for the higher Landau levels has
a qualitative dependence on the range of the disorder, showing
spatial oscillations for short-range disorder and becoming
qualitatively similar to the lowest Landau level situation only when
the disorder range exceeds the spatial extent of the excited
wavefunction. Without a real theory for the localization transition,
it is difficult to quantify these qualitative considerations. We
believe, however, that disorder invariably mixes Landau levels and the
independent Landau level approximation breaks down, making the
localization transition universal even for short-range disorder as
demonstrated by our numerical results  in the presence of Landau level
coupling effects.

Before concluding we discuss a number of experimental strong-field
localization issues which are not understood currently. While
experiment seems to be in ``agreement'' with theory that the
localization exponent $\nu\simeq 2.3$, it is hard to understand why a
{\bf non-interacting} theoretical calculation of the localization
exponent should agree with the measured experimental value because, in
general, electron-electron interaction is known[24] to be a relevant
perturbation which changes the universality class. In this respect, it
is reassuring that the measured localization exponent($\kappa\approx
0.4$) for the {\bf fractional} quantum Hall transition[25] is the same
as the integer quantum Hall case. It is important to have some
theoretical idea about why interaction does not change the
universality class in the strong-field situation (or, if it does, why
the measured value of $\kappa$ is the same for integer and fractional
situations in apparent agreement with the non-interacting calculation
of the localization exponent $\nu$). The experimentally found
dependence[26] of the measured $\kappa$ on the spin degeneracy of the
Landau level is also not understood at the present time. In
particular, $\kappa$ for the spin-unpolarized case is found[26] to be
half($\approx 0.21$) the value of the spin-polarized
situation($\approx 0.43$). In a non-interacting localization model, it
is unclear how the electron spin can be a relevant perturbation.
Inclusion of interaction effects, however, leads to a spin dependence
of fermion localization properties in the zero-field case[27]. Whether
the observed spin dependence of the exponent $\kappa$ is somehow
related to interaction effects or not is totally unknown at this
stage. We suggest a simple scenario for the spin dependence of
$\kappa$ which can be experimentally tested. Suppose that there is a
small (but unresolved) spin-splitting $\Delta E$ in the spin
degenerate unpolarized case. In that case, it is possible that, if
$\Delta E<<k_BT$, the experimental measurement of $\kappa$ measures
only an effective $\kappa$ which is smaller than the real $\kappa$
because there are two unresolved critical energies separated by a
small (but finite) $\Delta E$ and one is observing the combined
effects of both the localization transitions without resolving them.
(The argument fails if $\Delta E=0$, but the spin-splitting is
unlikely to be exactly zero.) A simple analysis shows that this
scenario leads to a temperature dependent effective $\kappa$ which
asymptotically approaches the real $\kappa (\approx 0.4)$ in the high
temperature limit but is lower than the real $\kappa$ in the
experimentally feasible regime, approaching, in fact, a
saturation({\it i.e.}$\kappa =0$) logarithmically in the low
temperature limit. Thus, a measurement of this effective $\kappa$ in
an intermediate temperature range could produce $\kappa\approx 0.2$ as
observed experimentally. This proposed scenario should be tested
experimentally through detailed temperature dependent measurements in
the spin degenerate situation. Finally, we mention that very recent
microwave frequency-dependent measurements[28] show clear evidence for
finite-frequency($f$) dynamic scaling of the strong-field localization
transition in the integer quantum Hall regime. The experimental
finding is that for $f\simgt 1GHz$, $\sigma_{xx}$ peaks broaden (at
fixed temperature) with increasing frequency, roughly as $\Delta B\sim
f^{\gamma}$ where $\gamma\approx 0.4(0.2)$ for
spin-split (spin-degenerate) peaks. Since one expects $\gamma=1/\nu z$
where $z$ is the dynamical exponent, one concludes that $z=1$ and
$\nu =2.3(4.6)$ for spin-split (spin-degenerate) situations. The
spin-dependence of the microwave experiments is, of course, consistent
with that of the temperature dependent measurements and remains
unexplained (unless our ``trivial'' explanation applies!). The
interesting finding of $z=1$ is consistent with the recent
speculation[29] for dirty boson systems in the context of
superconductor-insulator transition in thin metal films.
Statistical transmutation properties of two dimensional quantum
systems allow mapping between fermion and boson systems, and it is,
therefore, tempting to speculate[29] that $z=1$ is a universal feature
of all two dimensional quantum phase transitions[30].
 One problem in this context
is that theory[29] predicts a universal value of the critical
resistance at the metal-insulator transition whereas
experimentally[31] the value of the $\sigma_{xx}$ peak seems to be
non-universal and substantially below the universal
theoretical value of $e^2/h$. One should, however, bear in mind that
while the theory applies strictly at $T=0$, the experimental
measurements are necessarily at finite temperature, and a simple
extrapolation to $T=0$ to obtain the critical conductance
 may not work. This issue also requires further
investigation.

Our conclusion is
that the strong-field two-dimensional Landau level localization is indeed
universal({\it i.e.} $\nu_0=\nu_1$, {\it etc.}) except for the
short-range random disorder potential which, in the
absence of any Landau level coupling, gives rise to non-universal
localization with different Landau levels having different
localization exponents. Our work indicates, however, that inclusion of
a finite potential range and/or coupling between Landau levels
restores localization universality making $\nu_0=\nu_1$.

\begin{center}
{\bf ACKNOWLEDGEMENTS}
\end{center}

 This work is supported by the National Science Foundation.
The authors thank
Dr. Song He and Prof. X.C. Xie for helpful discussions and
suggestions. Part of our calculation was carried out on UMIACS CM2 and CM5
Connection Machines at the University of Maryland, College Park.

\pagebreak


\begin{center}
{\Large \bf References}
\end{center}

[1] {\em The Quantum Hall Effect}, edited by R.E.Prange
and S.M.Girvin (Springer-Verlag, New York, 1990).

[2] E.Abrahams, P.W.Anderson, D.C.Licciardello, and T.V.Ramakrishnan,
Phys.Rev.Lett. {\bf 42}, 673 (1979).

[3] H.P.Wei, D.C.Tsui, M.Paalanen, and A.M.M.Pruisken,
Phys.Rev.Lett. {\bf 61}, 1294 (1988); H.P.Wei, S.W.Hwang, D.C.Tsui,
and A.M.M.Pruisken, Surf.Sci. {\bf 229}, 34 (1990).

[4] J.Wakabayashi, M.Yamane, and S.Kawaji, J.Phys.Soc.Jpn. {\bf 58},
1903 (1989).

[5] S.Koch, R.J.Haug, K.von Klitzing, and K.Ploog, Phys.Rev.B {\bf
43}, 6828 (1991) and Phys.Rev.Lett. {\bf 67}, 883 (1991).

[6] V.T.Dolgopolov, A.A.Shaskin, B.K.Medvedev, and V.G.Mokerov, JETP
{\bf 72}, 113 (1991).

[7] M.D'Iorio, V.M.Pudalov, and S.G.Semenchinsky, {\em High Magnetic Fields
in Semiconductor Physics III}, edited by G. Landwehr
(Springer-Verlag, Berlin, 1992), P.56.

[8] S.Koch, R.J.Haug, K.von Klitzing and K.Ploog, Mod.Phys.Lett.B {\bf
6}, 1 (1992); Phys.Rev.B {\bf 46}, 1596 (1992).

[9] H.P.Wei, S.Y.Lin, D.C.Tsui, and A.M.M.Pruisken, Phys.Rev.B {\bf
45}, 3926 (1992).

[10] B.Huckstein and B.Kramer, Phys.Rev.Lett. {\bf 64}, 1437 (1990)
and references therein.

[11] H.Aoki and T.Ando, Phys.Rev.Lett. {\bf 54}, 831 (1985); T.Ando
and H. Aoki, J.Phys.Soc.Jpn. {\bf 54},2238(1985).

[12] B. Mieck, Europhys.Lett. {\bf 13}, 453 (1990).

[13] Y.Huo and R.N.Bhatt, Phys.Rev.Lett. {\bf 68}, 1375 (1992).

[14] A.M.M.Pruisken, Phys.Rev.Lett. {\bf 61}, 1297 (1988).

[15] S.Das Sarma and D.Liu, preprint(1993) and references therein.

[16] J.A.Simmons, H.P.Wei, L.W.Engel, D.C.Tsui, and M.Shayegan,
Phys.Rev.Lett. {\bf 63}, 1731 (1989).

[17] D.Liu and S.Das Sarma, Mod.Phys.Letter.B {\bf 7}, 449 (1993); and
p.971, Proceedings of the 21st ICPS (World Scientific, Singapore, 1992).

[18] T.Ando and Y.Uemura, J.Phys.Soc.Jpn. {\bf 36}, 959 (1974).

[19] G.Benettin, L.Gagani, A.Giorgilli, and J.M.Streley, Meccanica
{\bf 15}, 9 and 21 (1980); S. Das Sarma, S.He, and X.C.Xie,
Phys.Rev.Lett. {\bf 61}, 2144 (1988), and Phys.Rev.B {\bf 41}, 5544 (1990).

[20] H. Aoki, J.Phys.C {\bf 10}, 2583(1977).

[21] S.A.Trugman, Phys.Rev.B {\bf 27}, 7539 (1983).

[22] J.T.Chalker and P.D.Coddington, J.Phys.C {\bf 21}, 2665 (1988).

[23] G.V.Mil'nikov and I.M.Sokolov, JETP lett. {\bf 48}, 536 (1988).

[24] T.R.Kirkpatrick and D.Belitz, Phys.Rev.B {\bf 41}, 11082 (1990);
and references therein.

[25] L.Engel, H.P.Wei, D.C.Tsui, and M.Shayegan, Surf.Sci. {\bf 229},
13 (1990).

[26] H.P.Wei, S.W.Hwang, D.C.Tsui, and A.M.M.Pruisken, Surf.Sci. {\bf
229}, 34 (1990).

[27] T.R.Kirkpatrick and D.Belitz, Phys.Rev.B {\bf 45}, 3187 (1992);
and references therein.

[28] L.W.Engel, D.Shahar, C.Kurdak, and D.C.Tsui, preprint(1993).

[29] M.P.A.Fisher, G.Grinstein, and S.M.Girvin, Phys.Rev.Lett. {\bf
64}, 587 (1990); M.P.A.Fisher, Phys.Rev.Lett. {\bf 65}, 923 (1990)

[30] Note that one can express the temperature dependence of the
transition regime $\Delta B$ defining the width of the peak resistance
at the (metal-insulator) transition between quantum Hall plateaus as
$\Delta B\propto T^{\kappa}$ with $\kappa\equiv 1/\nu z_T$ where $z_T$
is a thermal dynamical exponent. We have followed the standard
convention, as used in Refs.3,5,8 and 9 for example, in writing
$z_T=2/p$ where $p$ is the thermal scaling exponent for the inelastic
scattering rate. Experimentally, of course, $z_T\approx 1$(because
$\kappa\approx 0.4$ and $\nu\approx 2.3$) which makes $z_T$ the same
as the $z$ obtained from frequency scaling experiments. This may not
be a mere coincidence -- see Ref.29 and, also, M.P.A.Fisher,
P.B.Weichman, G.Grinstein, and D.S.Fisher, Phys.Rev.B {\bf 40}, 546 (1989).

[31] D.C.Tsui, private communication.

\pagebreak


\begin{center}
{\Large\bf Table and Figure Caption}
\end{center}

\begin{tabbing}
xxx\=xxxxxxxxxxxxxxxxxxxxxxxxxxxxxxxxxxxxxx\kill
 \> \\
 \> \parbox{16cm}{ {\bf Table 1.} Critical exponents for different Landau
levels without Landau level coupling for weak $\delta$-function
impurities. }\\
 \> \\
 \> \parbox{16cm}{ {\bf Table 2.} Critical exponents for weak
repulsive $\delta$-function  impurities without Landau level coupling.}\\
 \> \\
 \> \parbox{16cm}{ {\bf Table 3.} Critical exponents for weak
finite range Gaussian impurity-potentials without Landau level coupling.}\\
 \> \\
 \> \parbox{16cm}{{\bf Table 4.} Critical exponents for strong
$\delta$-function impurity-potentials considering coupling between $N=0$ and
$N=1$ Landau levels.}\\
 \> \\
 \> \parbox{16cm}{{\bf Figure 1.} $N=0$ Landau level, for weak
$\delta$-function  impurities($c_i=8$):
(a) The renormalized finite system localization length $\lambda_M/M$
as a function of system size $M$, different symbols
represent different energies; (b) One-parameter scaling function, with inset:
Localization length $\xi (E)$, $\nu =2.32$}\\
 \> \\
 \> \parbox{16cm}{{\bf Figure 2.} $N=1$ Landau level, for weak
$\delta$-function  impurities($c_{i}=4$):
(a) $\lambda_M/M$ vs. $M$, different symbols
represent different energies; (b) One-parameter scaling function, with inset:
Localization length $\xi (E)$, $\nu=5.95$}\\
 \> \\
 \> \parbox{16cm}{{\bf Figure 3.} In the presence of weak repulsive
$\delta$-function impurities, $N=0, c_{i}=4$, the scaling function for the
lower(a) and higher(b) energy branches of the Landau band.
Corresponding localization length
in the critical regime are shown in the insets for lower energy branch,
$E_{c}=-0.15,
\nu=3.08$, and for higher energy branch, $E_{c}=-0.156, \nu=2.05$.}\\
 \> \\
 \> \parbox{16cm}{{\bf Figure 4.} For weak finite range Gaussian
impurity-potentials,
$N=0$ Landau band, the scaling function with localization length in
the insets are presented for (a) $d/l_{c}=0.5, c_{i}=8, \nu=2.21$; (b)
$d/l_{c}=1.0, c_{i}=4, \nu=2.3$.}\\
 \> \\
 \> \parbox{16cm}{{\bf Figure 5.} For weak finite range Gaussian
impurity-potentials,
$N=1$ Landau band, the scaling function with localization length in
the insets are presented for (a) $d/l_{c}=0.5, c_{i}=2, \nu=3.6$; (b)
$d/l_{c}=1.0, c_{i}=4, \nu=2.51$.}\\
 \> \\
 \> \parbox{16cm}{{\bf Figure 6.} For strong $\delta$-function
impurities considering coupling between $N=0$ and $N=1$ Landau levels,
the scaling function are
presented for $c_{i}=8$ (a) $N=0$ lower energy branch; (b) $N=0$
higher energy branch; (c) $N=1$ lower energy branch. Corresponding
localization length in the critical regime are presented in the insets
of (a)
$E_{c}=0.029, \nu=1.55$; (b) $E_{c}=0.100, \nu=2.24$; (c)
$E_{c}=0.872, \nu=2.85$.}\\
 \> \\
 \> \parbox{16cm}{{\bf Figure 7.} Shows the spatial behavior of the
calculated matrix elements of the impurity potential for $N=0$(solid
line), $N=1$(dotted line), and $N=2$(dashed line) Landau levels, and,
for four different values of the potential range
$d/l_c=0$(a),1(b),2(c), and 4(d). The impurity is located at $x_i$,
and $l_o=\sqrt{2\pi}l_c$ where $l_c$ is the magnetic length.}
\end{tabbing}

\vspace{1cm}

{\bf Table 1.}
\vspace{3mm}

\begin{tabular}{||c|c|c|c|c||}    \hline\hline
$c_i$     &    $\nu_{0}$    &$E_{c}$  &  $\nu_{1}$   &$E_{c}$\\ \hline
4      &    $2.21\pm0.02$ &0 &  $5.9\pm0.1$ & 0 \\ \hline
8      &    $2.32\pm0.02$ &0 &  $5.4\pm0.2$ & 0 \\ \hline
16     &    $2.22\pm0.04$ &0 &  $5.5\pm0.1$ & 0 \\ \hline
32     &    $2.22\pm0.02$ &0 &              &   \\ \hline\hline
Ando \em{et al}& $\leq2$  &0 &  $\leq4$     & 0 \\ \hline\hline
Kramer \em{et al} & $2.34\pm0.04$ &0 & $\sim4$&0 \\ \hline\hline
\end{tabular}

\pagebreak
\vspace{1cm}
{\bf Table 2.}
\vspace{3mm}

\begin{tabular}{||c|c|c|c|c||} \hline\hline
$c_i$ &$\nu_{0l}$ & $E_{c}$ &$\nu_{0h}$ & $E_{c}$  \\ \hline
4 &3.08 &-0.150 &2.05 &-0.156   \\ \hline
8 & & &2.07 &-0.109 \\ \hline
16 & & &2.22 &-0.090  \\ \hline
32 & & &2.29 &-0.075  \\ \hline\hline
\end{tabular}

\vspace{1cm}
{\bf Table 3.}
\vspace{3mm}

\begin{tabular}{||c|c|c|c|c|c||} \hline\hline
$d/l_{c}$ &$c_i$ &$\nu_{0}$ &$E_{c}$ &$\nu_{1}$ &$E_{c}$\\ \hline
 0.5  &2 &1.86 &0 &3.60 & 0\\ \cline{2-6}
      &4 &2.21 &0 & 3.56 & 0\\ \cline{2-6}
      &8 &2.21 &0 & 4.00 & 0\\ \hline
 1.0  &2 &2.13 &0 &2.8  & 0\\ \cline{2-6}
      &4 &2.30 &0 & 2.51 & 0\\ \cline{2-6}
      & 8 &2.14 &0 & 2.8 & 0\\ \hline\hline
\end{tabular}

\vspace{1cm}
{\bf Table 4.}
\vspace{3mm}

\begin{tabular}{||c|c|c|c|c|c|c|c||} \hline\hline
$\gamma=\Gamma/\hbar\omega_{c}$ &$c_i$
& $\nu_{0l}$ & $E_{c}$
 & $\nu_{0h}$ & $E_{c}$ & $\nu_{1l}$ & $E_{c}$ \\ \hline
0.5 &2 & 0.86 & 0.009 & 2.02 & 0.059 & 3.65 & 0.975
\\ \cline{2-8}
    &4 & 1.50 & 0 & 2.30 & 0.082 & 3.04 & 0.886 \\
\cline{2-8}
    &8 & 1.55 & 0.029 & 2.24 & 0.100 & 2.85 & 0.872 \\
\cline{2-8}
   &16 & 1.39 & 0.003 & 2.06 & 0.105 & 2.77 & 0.841 \\ \hline\hline
\end{tabular}

\end{document}